\documentclass[acus]{pac99}

\usepackage{epsfig}

\newcommand{\bit}{\begin{Itemize}}
\newcommand{\eit}{\end{Itemize}}

\begin{document}
\title{STUDIES FOR MUON COLLIDER PARAMETERS AT CENTER-OF-MASS ENERGIES
OF 10 TEV AND 100 TEV}
\author{ Bruce J. King, Brookhaven National Laboratory $^1$ }
\maketitle

\begin{abstract}

 Parameter lists are presented for speculative muon colliders at
center-of-mass energies of 10 TeV and 100 TeV. The technological advances
required to achieve the given parameters are itemized and discussed, and
a discussion is given of the design goals and constraints. An important
constraint for multi-TeV muon colliders is the need to minimize neutrino
radiation from the collider ring.

\end{abstract}

\footnotetext[1]{
web page: http://pubweb.bnl.gov/people/bking/,
email: bking@bnl.gov.
This work was performed under the auspices of
the U.S. Department of Energy under contract no. DE-AC02-98CH10886. }

\section{Introduction}
\label{sec-intro}

 The main motivation for research and development efforts on muon collider
technology is the assertion that affordably priced muon colliders might
provide lepton-lepton collisions at much higher center of mass (CoM) energies
than is feasible
for electron colliders, and perhaps eventually explore the spectrum of
elementary particles at mass scales inaccessible even to hadron colliders.

 This paper attempts a first look at these assertions
through discussion and evaluation of the self-consistent muon
collider parameter sets given in table 1, at CoM energies of 10 TeV
and 100 TeV.
These parameter sets have the purpose of pin-pointing the challenges
of very high energy muon colliders and they have not been studied or
discussed in detail within the Muon Collider Collaboration (MCC) or
elsewhere.

 The 10 TeV parameter set was presented previously~[1]
and the parameter values appear to be internally consistent.
In contrast, the 100 TeV parameter set represents work in progress
to improve in luminosity and other parameters over the self-consistent
100 TeV parameter set given in reference~[1] and it's parameters
are not yet fully consistent, as discussed below.

\section{Generation of Parameter Sets}
\label{sec-how}

 As described previously~[1],
the parameter sets in table 1 were generated through iterative runs of a
stand-alone FORTRAN program, LUMCALC. The parameter sets are calculated
from the input values for several input parameters --
namely, the CoM energy (${\rm E_{CoM}}$), the collider ring circumference (C)
and depth below the
Earth's surface (D), the beam momentum spread ($\delta$) and 6-dimensional
invariant
emittance (${\rm \epsilon_{6N}}$), the reference pole-tip magnetic field
for the final focus quadrupoles (${\rm B_{4\sigma}}$), and the
time until the beams are dumped (${\rm t_D}$) -- and from the input of
maximum allowable
values for several other parameters -- namely, the bunch repetition
frequency (${\rm f_b}$),
the initial number of muons per bunch ($N_0$), the beam-beam tune disruption
parameter ($\Delta \nu$),
the beam divergence at the interaction point ($\sigma_\theta$), the
maximum aperture for the final focus quadrupoles (${\rm A_{\pm4\sigma}}$), and
maximum allowable neutrino radiation where the plane of the collider ring
cuts the Earth's surface.

 As a preliminary stage of calculation, LUMCALC makes any parameter
adjustments that may be required to satisfy the input constraints. These
are, in order:
1) reducing $\sigma_\theta$ to the limit
 imposed by ${\rm A_{\pm4\sigma}}$ (based on scaling to existing final focus
designs at 0.1 TeV and 4 TeV[1]),
2) reducing ${\rm N_0}$ to
 attain an acceptable $\Delta \nu$, and
3) reducing ${\rm f_b}$ until
 the neutrino radiation is acceptable.

\section{Discussion}
\label{sec-discuss}

  The physics motivation for each of the parameter sets in table 1 is
discussed in~[2]. Briefly, the number of $\mu\mu \rightarrow {\rm ee}$
events gives
a benchmark estimate of the discovery potential for elementary particles
at the full CoM energy of the collider, while the production of hypothesized
100 GeV Higgs particles indicates roughly how the colliders would perform in
studying physics at this fixed energy scale.

  Both parameter sets give exciting luminosities with good potential
to explore the physics processes at and below their respective CoM
energy scales.

The output luminosity may be derived in terms of the input parameters as:
\begin{eqnarray}
\cal{L}
{\rm   [cm^{-2}.s^{-1}] }
   & = & {\rm 2.11 \times 10^{33} \times H_B
                   \times (1-e^{-2t_D[\gamma \tau_\mu]}) }
                                                           \nonumber \\
   & &  \times {\rm  \frac{ f_b[s^{-1}] (N_0[10^{12}])^2 (E_{CoM}[TeV])^3}
                          { C[km] } } \nonumber \\
   & & {\rm \times \left( \frac{\sigma_\theta [mr].\delta[10^{-3}]}
                   {\epsilon_{6N}[10^{-12}]}  \right) ^{2/3}  }.
                                    \label{lum}
\end{eqnarray}
This formula uses the standard MCC assumption[1] that
the ratio of transverse to longitudinal emittances can be chosen
in the muon cooling channel to maximize the luminosity for a
given ${\rm \epsilon_{6N}}$. The pinch enhancement factor, ${\rm H_B}$,
is very close to unity (see table 1), and the numerical coefficient
in equation 1 includes a geometric correction factor of 0.76 for the
non-zero bunch length,
$\sigma_z = \beta^*$ (the ``hourglass effect'') .

 In practice, the muon beam power and current are limited so the
optimization of parameters actually involves the ``specific luminosity''
\begin{equation}
l \equiv \frac{\cal{L}}{f_b \times N_0}.  \label{speclum}
\end{equation}
Further, the parameter sets at these high energies are always limited
by $\Delta \nu$ and it is useful and easy to rewrite equations~\ref{lum}
and~\ref{speclum} in the form
\begin{equation}
l \propto \frac{\Delta \nu}{\beta*},
\end{equation}
with no explicit dependence on emittance for a given energy.
This is misleading, however, since smaller emittances
mean smaller magnet apertures and hence allow the design of lattices
with smaller $\beta*$'s and, in practice, one can almost always
gain in specific luminosity by reoptimizing parameter sets with
smaller emittances.

  The assumed 6-dimensional emittances are factors of 3.5 (10 TeV)
and 50 (100 TeV) smaller than the value $170 \times 10^{-12}\;{\rm m}^3$
that is normally used in MCC scenarios for first generation muon
colliders, in anticipation that the muon cooling channel may eventually
be improved through further design optimization, stronger magnets,
higher gradient rf cavities and other technological advancements
and innovations.

  The final focus region presumably presents the most difficult
design challenge that is relatively specific to high energy muon
colliders. (The muon cooling channel is, of course, a formidable
challenge for all muon colliders.) Progressively stronger focusing
is required at higher energies to generate the smaller spot sizes
necessary to increase the luminosity. The strength of the
focusing can be gauged from the overall beam demagnification, defined as
$M \equiv \sqrt{\beta_{\rm max}/\beta^*}$. This is
a dimensionless parameter that should be closely correlated with
fractional tolerances in magnet uniformity, residual chromaticity
etc. Hence, it might be prudent to decrease the fractional momentum
spread of the beams, $\delta$, to compensate for an increasing $M$.
In the absence of final focus designs for these parameter sets
the difficulty of the chromatic correction for the final focus
has simplistically been assessed by the value of a single parameter:
the ``chromaticity quality factor'' $q \equiv M \delta$~[3].

 In the absence of ``proof by example'' lattice designs, the
next generation of proposed linear e+e- colliders (LC's) may well provide
the best benchmarks for heuristically assessing the difficulty of the
final focus parameters.
For the 10 TeV parameter set, each of $\beta*$, $M$ and $q$ are
roughly comparable to those proposed for these LC's. The 100 TeV
parameter set necessarily has a much larger $\beta*$ and $M$ but
the value of $\delta$ was forced down to $\delta=8 \times 10^{-5}$
to nearly hold steady the value of $q$. Unfortunately, this is
clearly inconsistent with the rising energy loss due to beamstrahlung
at the ip. Following the lead of the LC's, the next iteration of the
100 TeV parameter set will use colliding {\em flat beams} to make
the parameter sets fully consistent. The spot size -- clearly indicative of
vibration and alignment tolerances -- also falls with energy, but even at
100 TeV it remains an order of magnitude above the spot size in the
y coordinate for future LC design parameters.

  Neutrino radiation is an extremely serious problem for many-TeV
muon colliders and further information and discussion on this
can be found in these proceedings~[4].
 The radiation levels
are predicted to rise rapidly with the collider energy~[4]
and beyond CoM energies of a few TeV it may well be necessary to build
the colliders at isolated sites where the public would not be exposed to
the neutrino radiation disk.

\section{Conclusions}
\label{sec-conc}

 Muon colliders from 10 to 100 TeV CoM energies may offer an exciting
long-term future to experimental HEP and are not obviously infeasible
to this author. However, they clearly introduce some daunting new
challenges beyond those common to all muon colliders and the parameter
sets in table 1 will require some technological extrapolations.


\section{references}

\noindent [1] The Muon Collider Collaboration,
``Status of Muon Collider Research
and Development and Future Plans'', to be submitted to Phys. Rev. E.

\noindent [2] B.J. King,
 ``Discussion on Muon Collider Parameters at Center of Mass
Energies from 0.1 TeV to 100 TeV'', 19 June, 1998, 
     Submitted to Proc. Sixth European Particle Accelerator
Conference (EPAC'98), Stockholm, Sweden, 22-26 June, 1998.
Available at http://pubweb.bnl.gov/people/bking/.

\noindent [3] The idea to use this parameter emerged through
discussions with Frank Zimmermann and Carol Johnstone .

\noindent [4] B.J. King, ``Potential Hazards from Neutrino Radiation
at Muon Colliders'', these proceedings.

\newpage

\begin{table}[htb!]
\begin{center}
\caption{Example parameter sets for 10 TeV and 100 TeV muon colliders.
The generation of
these parameter sets is discussed in the text.
These parameters represent speculation by the author on how muon colliders
might evolve with energy. The beam parameters at the interaction point are
defined to be equal in the horizontal (x) and vertical (y) transverse
coordinates.}
\begin{tabular}{|r|cc|}
\hline
\multicolumn{1}{|c|}{ {\bf center of mass energy, ${\rm E_{CoM}}$} }
                             &  10 TeV  & 100 TeV \\
\hline \hline
\multicolumn{1}{|l|}{\bf collider physics parameters:} & & \\
luminosity, ${\cal L}$ [${\rm cm^{-2}.s^{-1}}$]
                                        & $1.0 \times 10^{36}$
                                        & $3.1 \times 10^{37}$ \\
$\int {\cal L}$dt [${\rm fb^{-1}/det/year}$]
                                        & 10 000
                                        & 310 000 \\
No. of $\mu\mu \rightarrow {\rm ee}$ events/det/year
                                        & 8700 & 2700 \\
No. of 100 GeV SM Higgs/det/year
                                         & $1.4 \times 10^7$
                                         & $6.5 \times 10^8$ \\
fract. CoM energy spread, ${\rm \sigma_E/E}$ [$10^{-3}$]
                                        & 1.0 & 0.08 \\
\hline
\multicolumn{1}{|l|}{\bf collider ring parameters:}  & & \\
circumference, C [km]                  & 15 & 100 \\
ave. bending B field [T]               & 7.0 & 10.5 \\
\hline
\multicolumn{1}{|l|}{\bf beam parameters:}            & & \\
($\mu^-$ or) $\mu^+$/bunch,${\rm N_0[10^{12}}]$
                                        & 2.4 & 2.0 \\
($\mu^-$ or) $\mu^+$ bunch rep. rate, ${\rm f_b}$ [Hz]
                                        & 15 & 10 \\
6-dim. norm. emittance, $\epsilon_{6N}
               [10^{-12}{\rm m}^3$]    & 50 & 3.5 \\
x,y emit. (unnorm.)
              [${\rm \pi.\mu m.mrad}$] & 0.55 & 0.0046 \\
x,y normalized emit.
              [${\rm \pi.mm.mrad}$]    & 26 & 22 \\
fract. mom. spread, $\delta$ [$10^{-3}$]
                                       & 1.4 & 0.084 \\
relativistic $\gamma$ factor, ${\rm E_\mu/m_\mu}$
                                        & 47 322
                                        & 473 220 \\
ave. current [mA]                      & 24 & 7.9 \\
beam power [MW]                        & 58 & 320 \\
decay power into magnet liner [kW/m]   & 1.4 & 2.4 \\
time to beam dump,
          ${\rm t_D} [\gamma \tau_\mu]$ & no dump & 0.5 \\
effective turns/bunch                  & 1039 & 985 \\
\hline
\multicolumn{1}{|l|}{\bf interaction point parameters:}      & & \\
spot size, $\sigma_x = \sigma_y
                          [{\rm nm}]$   & 780 & 93 \\
bunch length, $\sigma_z$ [mm]          & 1.1 & 0.185 \\
$\beta^*$ [mm]                          & 1.1 & 0.185 \\
ang. divergence, $\sigma_\theta$
                             [mrad]    & 0.71 & 0.5 \\
beam-beam tune disruption parameter, $\Delta \nu$
                                        & 0.100
                                        & 0.100 \\
pinch enhancement factor, ${\rm H_B}$   & 1.108
                                        & 1.130 \\
beamstrahlung fract. E loss/collision
                                       & $2.3 \times 10^{-7}$
                                       & $6.5 \times 10^{-4}$ \\
\hline
\multicolumn{1}{|l|}{\bf final focus lattice parameters:} & & \\
max. poletip field of quads., ${\rm B_{4\sigma}}$ [T]
                                        & 15 & 20 \\
max. full aperture of quad., ${\rm A_{\pm4\sigma}}$[cm]
                                        & 20 & 88 \\
${\rm \beta_{max} [km]}$               & 1100
                                        & 260 000 \\
final focus demagnification, $M \equiv \sqrt{\beta_{\rm max}/\beta^*}$
                                       & 31 000
                                       & $1.2 \times 10^6$ \\
chrom. quality factor, $Q \equiv M\delta$
                                       & 43 & 100 \\
\hline
\multicolumn{1}{|l|}{\bf synchrotron radiation parameters:} & & \\
syn. E loss/turn [MeV]                 & 17 & 25 000 \\
syn. rad. power [MW]                   & 0.4 & 200 \\
syn. critical E [keV]                  & 12 & 1700 \\
\hline
\multicolumn{1}{|l|}{\bf neutrino radiation parameters:} & & \\
collider reference depth, D[m]           & 300 & 300 \\
$\nu$ beam distance to surface [km]    & 62 & 62 \\
$\nu$ beam radius at surface [m]       & 1.3 & 0.13 \\
str. sect. length for 10x ave. rad., ${\rm L_{x10}}$[m] & 1.0 & 2.4 \\
ave. rad. dose in plane [mSv/yr]
                                        & 0.66 & 12.6 \\
\hline

\end{tabular}
\label{specs}
\end{center}
\end{table}

\end{document}